\documentclass[a4paper,11pt]{article}
\usepackage{jheppub} \usepackage{lineno}
\usepackage{besphysics}
\usepackage{subfigure}
\usepackage{comment}
\def \ee   {e^+e^-}
\def \uu   {\mu^+\mu^-}

\def \jpsi {J/\psi}
\def\cm{\ifmmode  {\mathrm{\ cm}}\else
                   \textrm{~cm}\fi}
\newcommand{\BESIIIorcid}[1]{\href{https://orcid.org/#1}{\hspace*{0.1em}\raisebox{-0.45ex}{\includegraphics[width=1em]{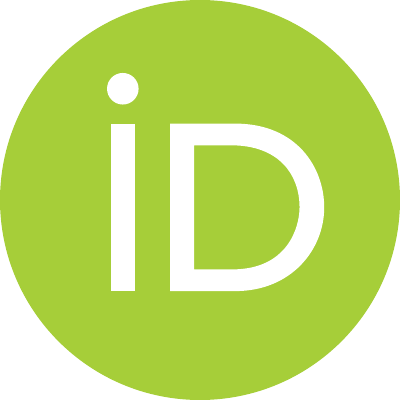}}}}

\collaborationImg{\includegraphics[width=0.15\textwidth, angle=90]{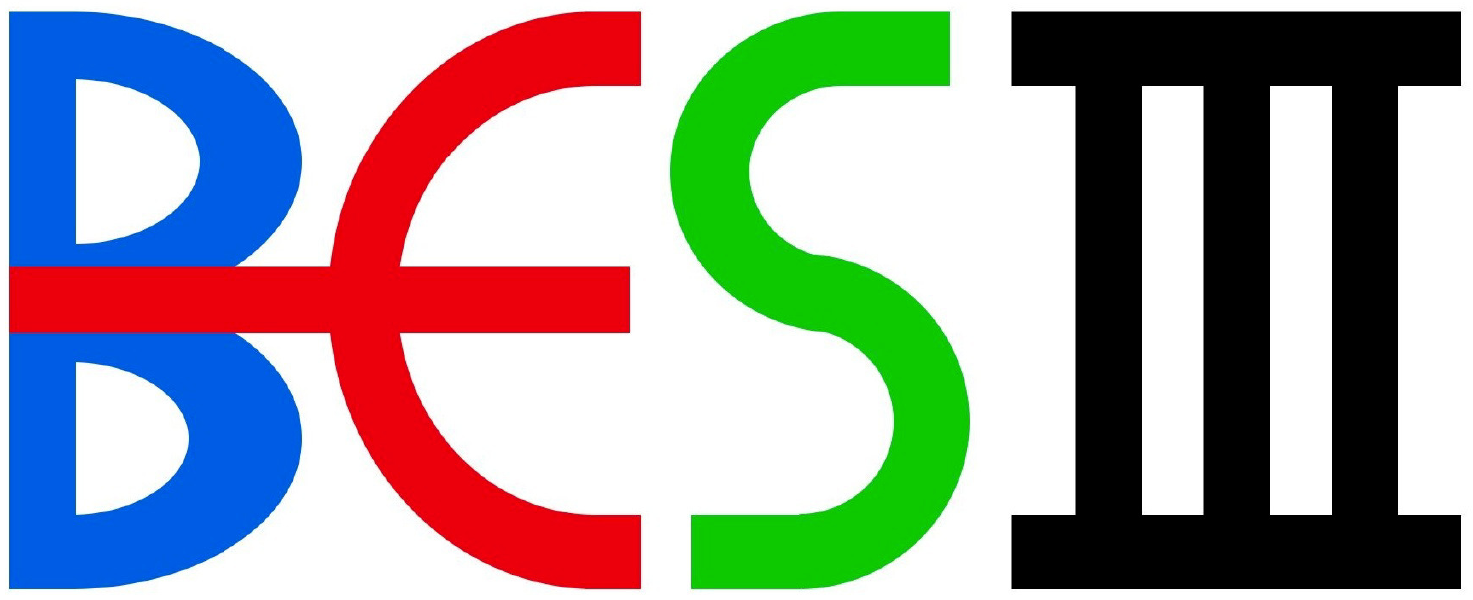}}
\collaboration{BESIII Collaboration}

\title{\boldmath Precision measurement of the $\jpsi\to\mu^+\mu^-\mu^+\mu^-$ branching fraction and search for a new gauge boson $Z^{\prime}$ at BESIII}

\abstract{We present a precision measurement of the branching fraction of the rare decay $\jpsi\to\uu\uu$, and a search for a new gauge boson $Z^{\prime}$ that couples only to the second and third lepton generations, using a data sample of $(8998 \pm 39) \times 10^6$ $\jpsi$ events collected by the BESIII detector.  The branching fraction of the $\jpsi\to\uu\uu$ decay is measured to be ${\cal B}(\jpsi\to\uu\uu)=[0.91\pm0.04~{\rm (stat.)}\pm0.07~{\rm (syst.)}]\times 10^{-6}$, representing the most precise measurement to date and consistent with the theoretical prediction within one standard deviation. 
Additionally, we search for $CP$ asymmetry in the $\jpsi\to\uu\uu$ decay and determine the T-odd observable to be $A_T = [0.05\pm 0.04~{\rm (stat.)}\pm0.01~{\rm (syst.)}]$, which is consistent with $CP$ conservation. In the search for $\jpsi\to Z^{\prime}\uu$ via $Z^{\prime}\to\uu$, no significant $Z^{\prime}$ signal is observed in the $\uu$ invariant mass region of $(0.212,~2.6)$~GeV$/c^2$, and we set upper limits on the ${\rm U}(1)_{L_{\mu}-L_{\tau}}$ model coupling constant $g^{\prime}$ ranging from 0.001 to 0.294 at the 90\% confidence level, which is comparable to the upper limits set by Belle II for $Z^{\prime}$ masses below $0.7$~GeV$/c^2$.}


\begin{document}
\maketitle
\flushbottom

\section{\boldmath Introduction}\label{sec:introduction}
\vspace{-0.3cm}

The branching fractions of purely leptonic decays of neutral vector mesons can be predicted precisely within the Standard Model (SM) since they do not involve large uncertainties from nonperturbative effects~\cite{Chen:2020bju,BESIII:jpsi4l}. Hence, the corresponding experimental measurements are particularly sensitive tests of the SM and provide opportunities to search for new physics. To date, the branching fractions of dilepton vector meson decays, $V \to \ell^{+}\ell^{-}$, have been measured quite precisely~\cite{pdg}. Here $V$ denotes a vector meson (such as $\rho^0$, $\omega$, $\phi$ and $J/\psi$) and $\ell^{\pm}$ denotes a light lepton, \emph{i.e.} $e^{\pm}$ or $\mu^{\pm}$. These measurements are important for the calculation of the nonperturbative decay constants $f_V$. Similar decay channels $J/\psi \to \ell^{+}\ell^{-}\ell^{\prime +}\ell^{\prime -}$, which provide stringent tests of Quantum Electrodynamics (QED) predictions, have been measured by the BESIII \cite{BESIII:jpsi4l}, CMS \cite{CMS:2024xwi} and LHCb\cite{LHCb:jpsi4l} experiments. The corresponding SM prediction is calculated at the leading order using QED amplitudes combined with non-relativistic Quantum Chromodynamics factorization \cite{Chen:2020bju}.

The rare decay $J/\psi\to\mu^+\mu^-\mu^+\mu^-$ has been recently observed by the CMS~\cite{CMS:2024xwi} and LHCb~\cite{LHCb:jpsi4l} collaborations, and the branching fraction results agree within 1.4 standard deviations with the leading-order QED calculation \cite{Chen:2020bju}.
The BESIII collaboration conducted a study investigating the decays $\jpsi\to\ee\ee$, $\jpsi\to\ee\uu$ and $\jpsi\to \uu\uu$, where the $\jpsi$ was produced through $\psi(3686)\to\pi^+\pi^-\jpsi$, using a data sample of $448.1\times 10^6$ $\psi(3686)$ events~\cite{BESIII:jpsi4l}. The first two processes were observed for the first time, and the measured branching fractions are consistent with theoretical predictions \cite{Chen:2020bju}. 
However, no significant signal was found for the  $\jpsi\to \uu\uu$ decay, primarily due to the low $\jpsi$ production yield and the low detection efficiency at the $\psi(3686)$ production threshold. 
The precision on the branching fraction of the decay $\jpsi\to\uu\uu$ can be much improved by utilizing the world's largest $\jpsi$ sample collected by the BESIII experiment at the center-of-mass energy $\sqrt{s}=3.097$~GeV, which is about 58 times larger than the $\psi(3686)$ sample.

The role of the muon in new physics scenarios has received much attention in recent years.  This is largely due to updated measurements of the anomalous magnetic moment of the muon by the Fermilab Muon $g-2$ experiment~\cite{Muong-2:2025xyk}, which achieved a more than four times improved precision over the BNM E821 measurement~\cite{Muong-2:2006E821}.
The theoretical predictions have been updated by the Muon $g-2$ Theory Initiative. The latest evaluation of the muon anomalous magnetic moment in Ref.~\cite{Aliberti:2025beg}, which updates the 2020 White Paper~\cite{Aoyama:2020ynm}, is consistent with the current world-average experimental measurement.
However, the hadronic vacuum polarization contribution in the theoretical prediction is based on measurements of the $e^+e^-\to \rm hadrons$ cross section, for which discrepancies among the various $e^+e^-$ collider experiments still persist \cite{DiLuzio:2024sps,BESIII:2015equ,BaBar:2012bdw,CMD-2:2006gxt,Xiao:2017dqv}. 
% This outcome has further intensified the demand for unprecedented precision in both experimental measurements and theoretical calculations, so that the consistency between SM predictions and experimental measurements can be scrutinized at an even higher precision. 
These motivate further studies of possible new interactions involving muons. In particular, new gauge bosons that couple to muons can contribute to the anomalous magnetic moment and can be probed through complementary direct searches.

\begin{figure}[htbp]
	\centering
	\subfigure{
		\includegraphics[width=0.7\linewidth]{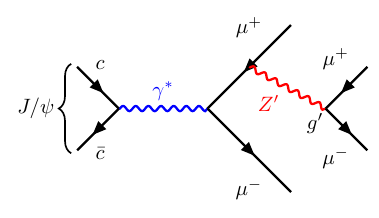}
	                }	
	\caption{A Feynman diagram for $J/\psi\to Z^{\prime}(\to\uu)\uu$ in a new physics scenario.}	
	\label{fig:Mumu_Z}
\end{figure}

One of the noted extensions of the SM gauge group is the anomaly-free gauged  ${\rm U}(1)_{L_{\mu}-L_{\tau}}$ model~\cite{He:1990pn, He:1991qd}. The  ${\rm U}(1)_{L_{\mu}-L_{\tau}}$ model introduces a new massive vector boson $Z^{\prime}$, which only couples to the second and third generations of leptons ($\mu,\nu_{\mu},\tau,\nu_{\tau}$) with the coupling constant $g^{\prime}$. Recent studies reveal that an extended ${\rm U}(1)_{L_{\mu}-L_{\tau}}$ model can also explain dark matter phenomena and the relic abundance of dark matter~\cite{Kamada:2018zxi,Foldenauer:2018zrz,Kahn:2018cqs,Patra:2016shz,Shuve:2014doa}. In the decay $\jpsi\to \uu Z^{\prime}$, a $Z^{\prime}$ could be radiated from one of the final state muons and then decay to another muon pair, as shown in figure \ref{fig:Mumu_Z}. Moreover, as Charge-Parity ($CP$) violation in the SM is too small to explain the observed matter-antimatter asymmetry in the Universe; new physics contributions, such as the $Z^{\prime}$, could introduce new $CP$ violation mechanisms \cite{Langacker:2000ju,Chang:2009wt}. Therefore, both the study of the $\uu$ invariant mass spectra  and the search for $CP$ violation in the $\jpsi\to \uu\uu$ decay could help constrain new physics contributions.

Experimentally, searches for the $Z^\prime$ decaying to a pair of muons 
have been reported by the BaBar~\cite{BaBar:2016sci}, CMS~\cite{CMS:2018yxg}, Belle~\cite{Belle:2021feg}, and Belle II~\cite{BelleII:gp} collaborations. With the wold's largest $\jpsi$ data set collected with BESIII, we have a chance to improve the constraint in the $Z^{\prime}$ parameter space.

This paper presents a branching fraction measurement for the rare electromagnetic decay $\jpsi\to\uu\uu$, followed by a search for $CP$ violation in the decay using the T-odd observable $A_T$. Finally, a search is performed for a hypothetical $Z^{\prime}$ boson via $\jpsi\to Z^{\prime}(\to\mu^+\mu^-)\mu^+\mu^-$ in the selected four-muon events.
The work is based on a data sample of 9 billion $\jpsi$ events collected by the BESIII detector in 2009, 2018, and 2019~\cite{lum-mer}.  The data collected in 2012 is not used due to the absence of information from the muon counter detectors.

\section{\boldmath BESIII detector and Monte Carlo simulation}\label{sec:BESIII Detector}
\vspace{-0.3cm}
The BESIII detector~\cite{Ablikim:2009aa} records symmetric $e^+e^-$ collisions 
provided by the BEPCII storage ring~\cite{Yu:2016cof}
in the center-of-mass energy range from 1.84 to 4.95~GeV,
with a peak luminosity of $1.1 \times 10^{33}\;\text{cm}^{-2}\text{s}^{-1}$ 
achieved at $\sqrt{s} = 3.773\;\text{GeV}$. 
The cylindrical core of the BESIII detector covers 93\% of the full solid angle and consists of a helium-based
 multilayer drift chamber, a time-of-flight
system~(TOF), and a CsI(Tl) electromagnetic calorimeter~(EMC),
which are all enclosed in a superconducting solenoidal magnet
providing a 1.0~T magnetic field.
The solenoid is surrounded by an octagonal flux-return yoke made of steel, interleaved with resistive-plate-counter muon-identification modules (MUC).
The charged-particle momentum resolution at $1~{\rm GeV}/c$ is
$0.5\%$, and the 
${\rm d}E/{\rm d}x$
resolution is $6\%$ for electrons
from Bhabha scattering. The EMC measures photon energies with a
resolution of $2.5\%$ ($5\%$) at $1$~GeV in the barrel (end cap)
region. The time resolution in the plastic scintillator TOF barrel region is 68~ps, while
that in the end cap region was 110~ps. {The end cap TOF
system was upgraded in 2015 using multigap resistive plate chamber
technology, providing a time resolution of
60~ps,
which benefits 97\% of the data used in this analysis~\cite{etof1,etof2,etof3}. 

Monte Carlo (MC) simulated data samples produced with a {\sc
geant4}-based~\cite{geant4} software package, which
includes the description of the BESIII detector geometry and response, are used to determine detection efficiencies
and to estimate backgrounds. The simulation models the beam
energy spread and initial state radiation in the $e^+e^-$
annihilations with the generator {\sc
kkmc}~\cite{ref:kkmc1,ref:kkmc2}.
The inclusive MC sample includes both the production of the $J/\psi$
resonance and the continuum processes incorporated in {\sc
kkmc}~\cite{ref:kkmc1,ref:kkmc2}.
All particle decays are modelled with {\sc
evtgen}~\cite{ref:evtgen1,ref:evtgen2} using branching fractions 
either taken from the
Particle Data Group ~\cite{pdg}, when available,
or otherwise estimated with {\sc lundcharm}~\cite{ref:lundcharm1,ref:lundcharm2}.
Final state radiation~
from charged final state particles is incorporated using the {\sc
photos} package~\cite{photos2}.
The signal process $\jpsi\to\uu \uu$ is generated based on the amplitudes in ref.~\cite{Chen:2020bju}. The events for $\jpsi\to\uu Z^{\prime}$ with $Z^{\prime}\to\uu$ are generated using a custom model based on ref.~\cite{Cvetic:2020vkk} for the $Z^{\prime}$ production, while the decay is modeled as a vector to a pair of leptons. Since the total width of the $Z^{\prime}$ is predicted to be much smaller than the detector resolution, it is assumed to be zero in the MC simulation~\cite{Rizzo:2006nw}. The two-photon processes $\ee\to\gamma^*\gamma^*\to\ee\uu$ and $\ee\to\gamma^*\gamma^*\to\uu\uu$ are simulated with {\sc diag36}~\cite{Berends:1984gf}. The decay  $\jpsi\to\ee\uu$ is simulated using the theoretical calculation~\cite{BESIII:jpsi4l}.

\section{\boldmath Event selection and background study}\label{sec:Event Selection and Background Study}
In this analysis, candidate signal events are reconstructed from four charged tracks with zero net charge. Each charged track is required to originate from the interaction point (IP) and the distance of closest approach to the IP must be less than 10\,cm along the $z$-axis, $|V_{z}|$, and less than 1\,cm in the transverse plane, $|V_{xy}|$. The absolute value of the cosine of the polar angle, $|\cos\theta|$, of each charged track is required to be less than 0.93, where $\theta$ is defined with respect to the $z$ axis, which is the symmetry axis of the multilayer drift chamber. To identify a track as a muon, the deposited energy in the EMC for each track [$E_{\rm EMC}(\mu)$] is required to fall between 0.1 and 0.3 GeV. 
Furthermore, the hit depth in the MUC for each track 
is required to be greater than either $\left[70\times |p_{\rm trk}|\left({\rm GeV}/c\right)^{-1}-40\right]$~cm for $0.5 \leq p \leq 1.1~{\rm GeV}/c$ or 40~cm for $p > 1.1~{\rm GeV}/c$, where $p_{\rm trk}$ is the track momentum.

According to theoretical predictions, signal events are mainly distributed in the low-$q^2$ region. The minimum invariant mass of any muon pair combination, ${\rm Min}\{M(\uu)\}$, is used to separate signal from background because of its narrow signal region compared to other possible choices. The dominant background therefore originates from low-momentum electrons and pions that are easily misidentified as muons.

Events in which a real photon converts into a low-momentum $\ee$ pair either in the beam pipe or in the inner wall of the drift chamber are a source of background when both electron tracks are misidentified as muons. To suppress this background, the $\uu$ pair with minimum $M(\uu)$ must have a reconstructed $x-y$ vertex outside of the region [2.0, 8.0]~cm.

% To suppress background originating from photon conversions, where a photon converts into a low momentum $\ee$ pair either in the beam pipe or in the inner wall of the drift chamber, the $\uu$ pair with minimum $M(\uu)$ must have a reconstructed $x-y$ vertex outside of the region [2.0, 8.0]~cm.

Furthermore, background from the $\jpsi\to\ee\uu$ process is significantly suppressed by imposing further event selection criteria on the two muons associated with ${\rm Min}\{M(\uu)\}$. Events where the absolute difference in momentum of the two muons, $|p(\mu^+)-p(\mu^-)|$, is less than 0.2 GeV/$c$ and where the cosine of their opening angle, $\cos(\uu)$, is greater than 0.995, are excluded. By applying this requirement, about 43.3\% of $\jpsi\to\ee\uu$ events are suppressed, at a cost of only 0.5\% of signal events.

The MC simulations for the two-photon processes $\ee\to\uu\uu$ and $\ee\to\ee\uu$ are validated by comparing control samples of data with MC samples at a center-of-mass energy of 3.773 GeV. Theoretically no interference between the signal process $\jpsi\to\uu\uu$ and the background process $\ee\to\uu\uu$ is expected. The simulation of $\ee\to\ee\uu$ is in good agreement with the data, while that of $\ee\to\uu\uu$ exhibits significant discrepancy with respect to the data. The difference between data and MC is defined as $\Delta d=\frac{N_{\rm MC}-N_{\rm Data}}{N_{\rm MC}}=0.45 \pm 0.07$, where $N_{\rm MC}$ and $N_{\rm Data}$ are the $\ee\to\uu\uu$ yields estimated from MC simulation and that of data, respectively. 
The large $\Delta d$ value, understood to be caused by the absence of initial state radiation corrections in the {\sc diag36} generator~\cite{Belle:2021feg}, is used to correct the $\ee\to\uu\uu$ events simulated at $\sqrt{s}=3.097$ GeV, 
and the corresponding uncertainty is considered as a source of systematic uncertainty.

To suppress background from pion misidentification, two four-constraint (4C) kinematic fits are performed that constrain the total reconstructed four-momentum of the four final-state particles to the initial four-momentum of the colliding beams, yielding fit $\chi_{\rm 4C}^2(4\pi)$ and $\chi_{\rm 4C}^2(4\mu)$ under the all-pion and all-muon hypotheses, respectively. The large background from the $\jpsi\to\pi^+\pi^-\pi^+\pi^-$ decay is effectively suppressed by requiring $\chi_{\rm 4C}^2(4\pi)>\chi_{\rm 4C}^2(4\mu)$
and $\chi_{\rm 4C}^2(4\mu)<14$. 
The residual $\jpsi\to\pi^+\pi^-\pi^+\pi^-$ background contribution is estimated from weighted MC events of $\jpsi\to\pi^+\pi^-\pi^+\pi^-$ to improve the agreement between data and simulation. The per-event weights are determined from a multivariate algorithm~\cite{Rogozhnikov:2016bdp} trained on the distributions of kinematic variables that include: $\chi_{\rm 4C}^2[4\pi(\mu)]$; the largest and second largest momenta of the four tracks; the momenta of the higher-momentum $\pi^+$ and $\pi^-$ tracks; the polar angles of the two $\pi^+$s and the lower-momentum $\pi^-$ tracks; the invariant mass of random combinations of $(\pi^+\pi^-)$; and the maximum and minimum invariant mass value of any $\pi^+\pi^-$ combination.  The kinematic variables are obtained from dedicated $\jpsi\to\pi^+\pi^-\pi^+\pi^-$ data and inclusive MC samples, which are produced with all the above selection criteria applied except for those on the MUC depth and $\chi_{\rm 4C}^2(4\pi)>\chi_{\rm 4C}^2(4\mu)$.

In addition to the various background processes described above, we use the inclusive MC sample to account for the remaining $J/\psi \to \text{others}$ decays, the dominant one being $J/\psi \to \omega(782) (\to \mu^+ \mu^-) \pi^+ \pi^-$.

The main background processes and their expected yields are summarized in table~\ref{tab:bkg}. They are used in the signal-extraction fit to constrain the fractions of the background components.

\begin{table}[htp]
	\centering
\begin{tabular}{c | c }
		\hline
		{\centering{Background}} &
		{\centering{Yield}}\\
		\hline

         $\jpsi\to\pi^+\pi^-\pi^+\pi^-$ & $143\pm 21$\\
         $\jpsi\to\ee\uu$ &$88\pm3$\\
		 $\ee\to\uu\uu$ & $115\pm 18$\\
		 $\ee\to\ee\uu$ & $3\pm2$\\
		 $\jpsi\to \text{others}$ & $17\pm 4$\\

 		\hline	
	\end{tabular}
    \caption{Major background sources and estimated yields.}
    \label{tab:bkg}
\end{table}

\section{\boldmath Branching fraction and $A_{T}$ measurement}\label{sec:Branching fraction and ACP measurement}
To extract the $\jpsi\to\uu\uu$ signal yield, an unbinned maximum likelihood fit is performed on the ${\rm Min}\{M(\uu)\}$ spectrum. The fit result is presented in figure \ref{fig:Fit_Minmpmm}. In this fit, the signal is described by a template shape constructed from signal MC simulation, while the background components are modeled with dedicated MC samples. 
The background yield of each component is a free parameter in the fit with a Gaussian constraint on the corresponding yields and uncertainties (as shown in table~\ref{tab:bkg}).

As shown in figure \ref{fig:Fit_Minmpmm}, we observe a signal from $\jpsi\to\uu\uu$ with a statistical significance exceeding 10$\sigma$. The significance is calculated by comparing the likelihoods with and without the signal components in the fit, taking the change of the number of degrees of freedom into account. The signal yield is $N_{\rm sig}=1030\pm 44$.

\begin{figure}[htbp]
	\centering
	\subfigure{
		\includegraphics[width=0.7\linewidth]{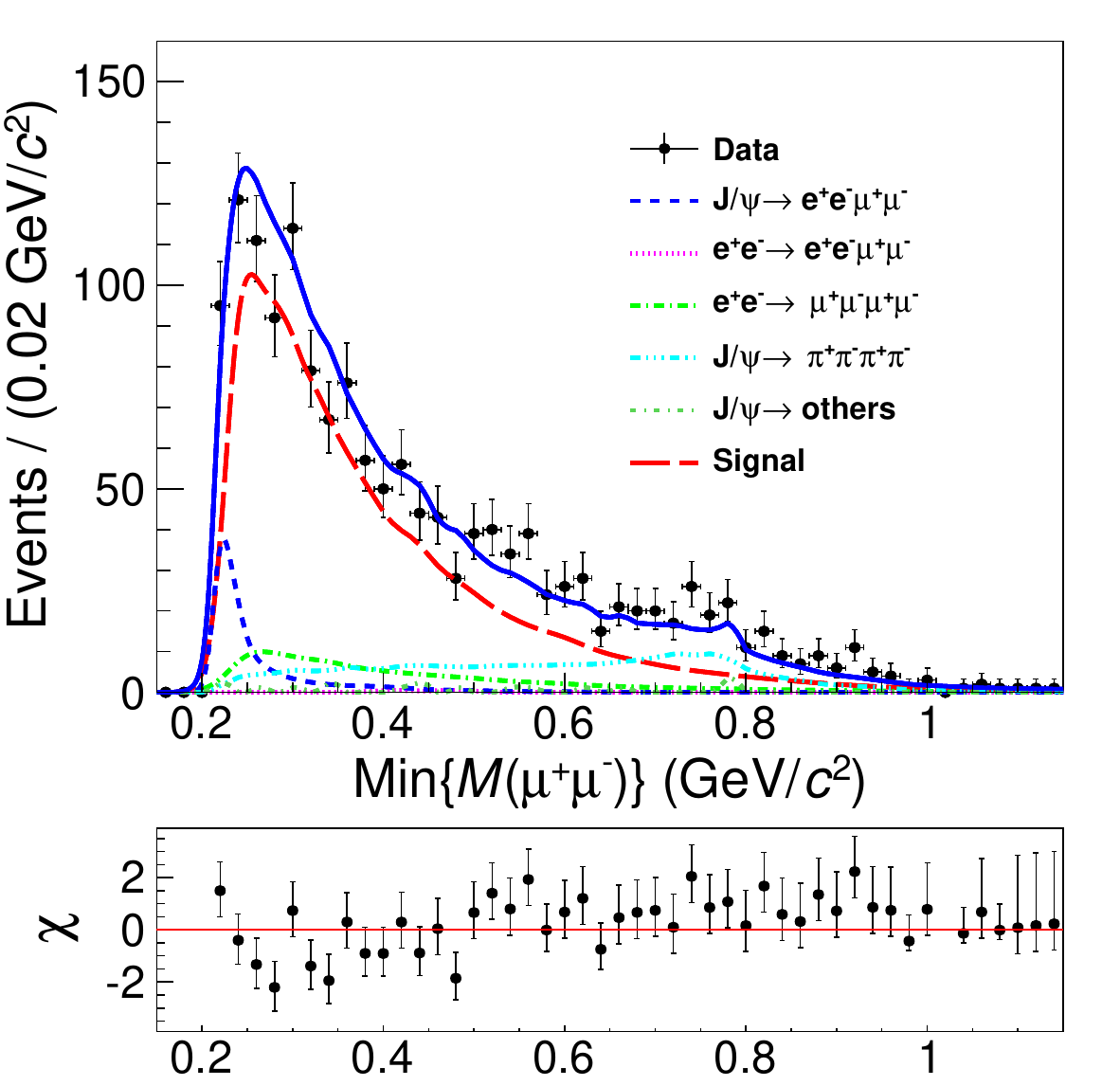}
	        }	
	\caption{The fit to the ${\rm Min}\{M(\uu)\}$ distribution from the selected data sample. Each fit component is presented in the legend. The normalized residual $\chi$ distribution is shown underneath the fit plot, where $\chi$ is defined as ($N_{\rm observed} - N_{\rm predicted}$)/$\sqrt{N_{\rm predicted}}$.}	
	\label{fig:Fit_Minmpmm}
\end{figure}

The branching fraction of $\jpsi\to\uu\uu$ is determined to be:
\begin{equation}
    \mathcal{B}(\jpsi\to\uu\uu)=\frac{N_{\rm sig}}{\epsilon \cdot N_{\jpsi}^{\rm tot}}= 
    \frac{1030\pm44}{0.126\times8998\times 10^{6}}=[0.91\pm0.04~{\rm(stat.)}]\times 10^{-6},
\end{equation}
where $N_{\rm sig}$, $\epsilon$, and $N_{\jpsi}^{\rm tot}$ are the signal yield, the detection efficiency, and the total number of $\jpsi$ events in data~\cite{BESIII:2021cxx}, respectively.
The resulting branching fraction is in good agreement with the theoretical prediction, $\mathcal{B}(\jpsi\to\uu\uu)=(0.974\pm 0.005)\times 10^{-6}$~\cite{Chen:2020bju}.

This analysis also searches for $CP$ asymmetry in $\jpsi\to\uu\uu$. 
A four-body decay gives rise to three independent final momenta in the rest frame of the decaying particle.  From these, one can form the T-odd observable ${C_T=(\Vec{p}_{l_{a}^+}\times\Vec{p}_{l_{a}^-})\cdot \Vec{p}_{l_{b}^+}}$~\cite{Gronau:2011cf, Bigi:2017eni}, where ${l_{a}^+,l_{a}^-}$ are the final state muons each with a lower momentum than its same-charge counterpart, $l_{b}^+$ is the remaining $\mu^+$, and all four-momenta are calculated in the $\jpsi$ rest frame. The $CP$ asymmetry observable based on Charge-Parity-Time invariance, $A_T$, can then be constructed from the numbers of events with positive and negative $C_T$ values as
\begin{equation}
    A_T=\frac{N(C_T>0)-N(C_T<0)}{N(C_T>0)+N(C_T<0)}.
\end{equation}
Using the same fit method described previously, two independent fits are performed for events with positive and negative $C_T$ values. The raw numbers of events with $C_T>0$ and $C_T<0$ are determined to be $554\pm29$ and $511\pm28$, while the corresponding signal efficiencies are 12.4$\%$ and 12.6$\%$, respectively.
Hence, after taking the corresponding signal efficiencies into account, $A_T$ is determined to be $0.05\pm 0.04$, where the uncertainty is statistical only.

\section{\boldmath Search for $Z^{\prime}$ in $\jpsi\to Z^{\prime}(\to\uu)\uu$}\label{sec:serarch}
In the search for the $Z^{\prime}\to\uu$ signal, all four possible combinations of the $\uu$ from $\jpsi\to\uu\uu$ are considered.
The selection criteria are the same as described above, except the requirements on $|p(\mu^+)-p(\mu^-)|$ and $\cos(\uu)$ are excluded since they lead to a significant drop in the $Z^{\prime}$ signal efficiency at masses near the di-muon mass threshold.

To extract the signal yield $N_{\rm sig}(Z^{\prime})$ as a function of the mass of the $Z^{\prime}$, $ M(Z^{\prime})$, we perform a series of unbinned maximum likelihood fits to the $M(\uu)$ spectrum. The search is conducted with a different $Z^{\prime}$ mass hypotheses in steps of the $Z^{\prime}$ mass resolutions. A total of 880 mass points have been probed in the mass range of $0.212<M(Z^{\prime})<2.6$~GeV$/c^2$. 
Figure~\ref{fig:nsig_scan} shows the extracted numbers of the signal events at different $M(Z^{\prime})$ hypotheses.
The region above $M(Z^{\prime})=2.6$~GeV$/c^2$ is not studied due to low efficiency and significant systematic uncertainty resulting from the high proportion of low momentum tracks.
In these fits, the probability density functions (PDFs) for all components are modeled as a template shape constructed from MC simulation. The $\jpsi\to\uu\uu$ process is treated as background and the yields of $\jpsi\to\uu\uu$ and $\jpsi\to Z^{\prime}\uu$ processes are free parameters in the fit. As shown in figure \ref{fig:735fit}, the largest upward local significance is determined to be 2.8$\sigma$ at $M(Z^{\prime})=0.735$~GeV with $N_{Z^{\prime}}^{\rm sig}=21\pm 9$.

\begin{figure}[htbp]
	\centering
	\subfigure{
		\includegraphics[width=0.9\linewidth]{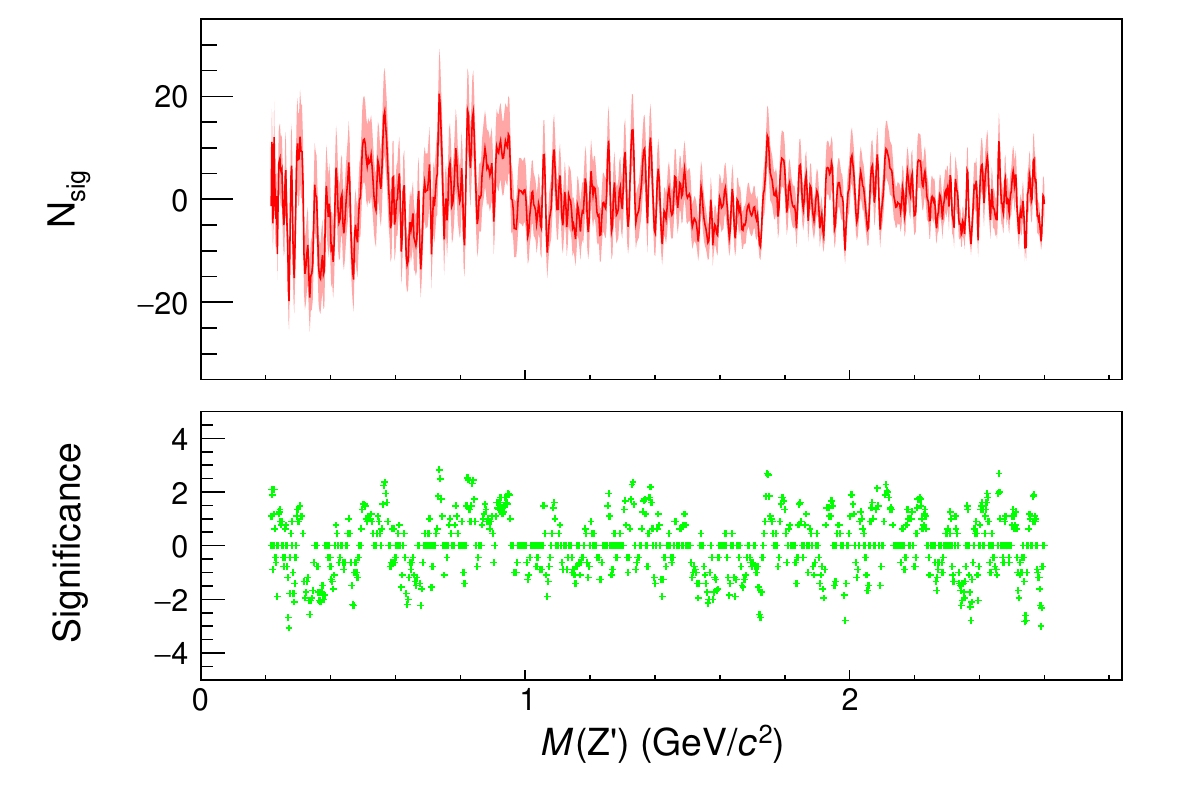}
	        }	
	\caption{The extracted numbers of signal events (red solid line) and the corresponding statistical uncertainties (red shaded region) at different $m_{Z'}$ hypotheses. The bottom plot shows the corresponding local signal significances (green dots).}	
	\label{fig:nsig_scan}
\end{figure}

\begin{figure}[htbp]
	\centering
	\subfigure{
		\includegraphics[width=0.7\linewidth]{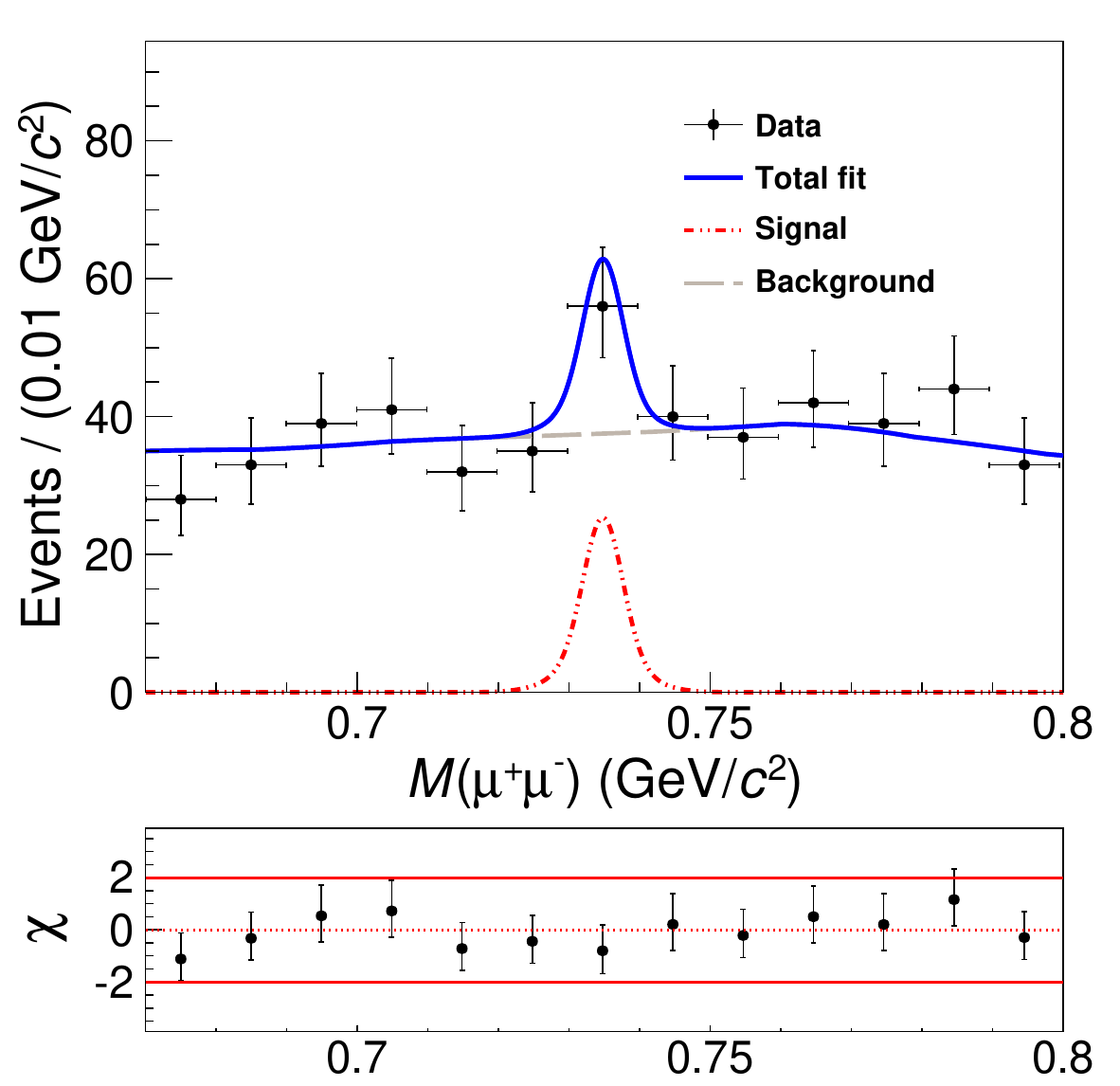}
	        }	
	\caption{The fit to the $M(\uu)$ distribution for $m_{Z'}=0.735$~GeV$/c^2$. Shown are the total fit (blue solid line), the $\jpsi\to Z^{\prime}\uu$ signal PDF (red dash-dot line), and the background PDF (grey dash line).}	
	\label{fig:735fit}
\end{figure}

Since no $Z^{\prime}$ signal is observed, upper limits on the product branching fractions at the 90$\%$ confidence level (CL) are calculated using $N_{\rm sig}^{\rm UL}$ obtained from a Bayesian method~\cite{Liu:2015uha} as:
\begin{equation}
 {\rm UL}_{\mathcal{B}(\jpsi\to Z^{\prime}\uu)\times \mathcal{B}(Z^{\prime}\to\uu)}=\frac{N_{\rm sig}^{\rm UL}}{\epsilon_{M(Z^{\prime})} N_{\jpsi}^{\rm tot}},
\end{equation}
where $N_{\rm sig}^{\rm UL}$ is the $Z^{\prime}$ signal yield at the 90$\%$ CL upper limit, and $\epsilon_{M(Z^{\prime})}$ is the detection efficiency at $M(Z^{\prime})$.

In addition, by considering the branching fraction $\mathcal{B}(Z^{\prime}\to\uu)$, the upper limit ${\rm UL}_{\mathcal{B}(\jpsi\to Z^{\prime}\uu)\times \mathcal{B}(Z^{\prime}\to\uu)}$ can be translated into constraints on the coupling constant
$g^{\prime}$ as~\cite{Araki:2017wyg,Cvetic:2020vkk}:
\begin{equation}
g^\prime = C[M(Z^\prime)]\sqrt{{\rm UL}_{\mathcal{B}(\jpsi\to Z^{\prime}\uu)\times \mathcal{B}(Z^{\prime}\to\uu)}}
\end{equation}
where $C[M(Z^\prime)]$ is a constant that depends only on $M(Z^\prime)$. The limits on the coupling constant $g^{\prime}$ range from 0.001 to 0.2942 in the range  $0.212<M(Z^{\prime})<2.6$~GeV$/c^2$, as shown in figure \ref{fig:coup_UL}. The previously excluded parameter spaces from BaBar~\cite{BaBar:2016sci}, Belle~\cite{Belle:2021feg} and Belle II~\cite{BelleII:gp} are also presented.

\begin{figure}[htbp]
	\centering
	\subfigure{
		\includegraphics[width=0.7\linewidth]{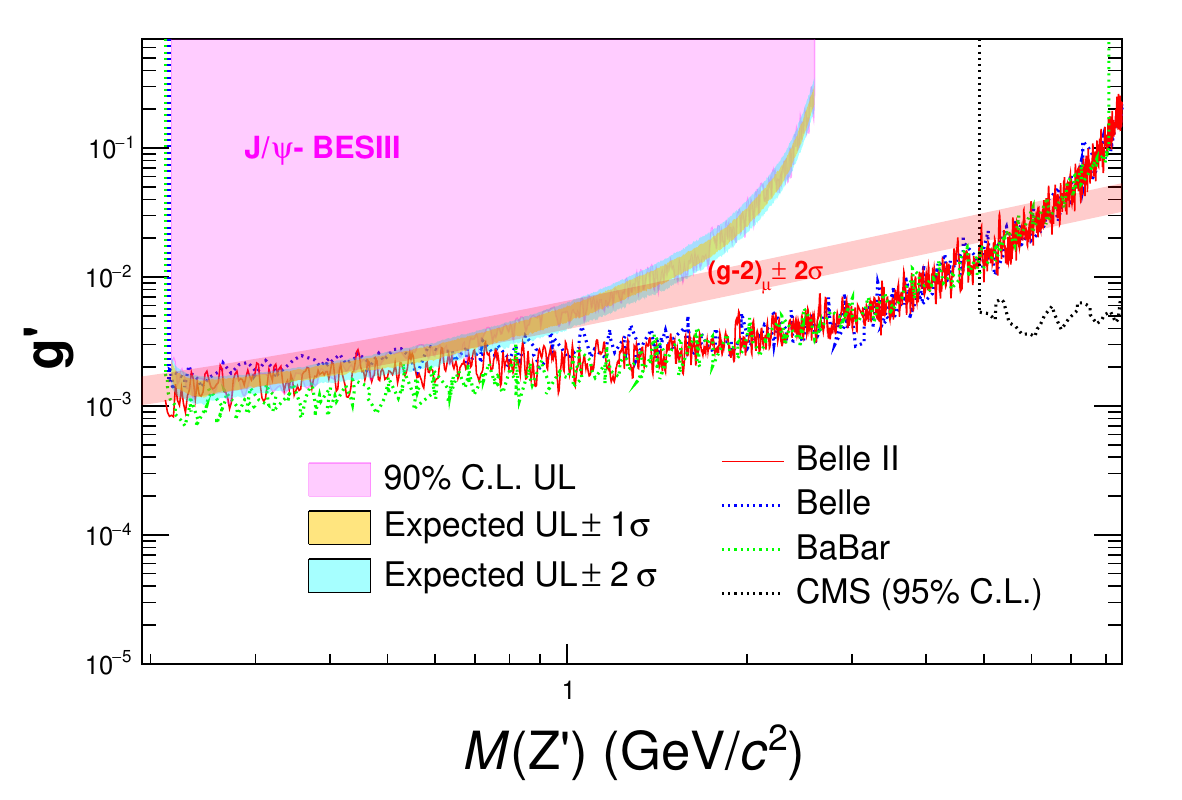}
	        }	
	\caption{The exclusion region at 90$\%$ CL for the coupling constant $g^{\prime}$ as a function of $M(Z^{\prime})$ (pink region), as well as the corresponding 68\% (inner yellow band) and 95\% (outer cyan band) expected limits. The red solid line represents the limits from Belle II~\cite{BelleII:gp}, the blue dotted line is from Belle~\cite{Belle:2021feg}, the green dotted line from BaBar~\cite{BaBar:2016sci} and the black dotted line from CMS~\cite{CMS:2018yxg} (95\% CL). The red band represents the values of the coupling constant $g^{\prime}$ needed to explain the $(g-2)_{\mu}$ anomaly within $2\sigma$, based on the analysis of the 2020 Whitepaper from the Theory Initiative.}
	\label{fig:coup_UL}
\end{figure} 

\section{\boldmath Systematic uncertainty}\label{Systematic Uncertainties}

Several sources of systematic uncertainty are considered in the measurements of $\mathcal{B}(\jpsi\to\uu\uu)$ and $\mathcal{B}(\jpsi\to Z^{\prime}\uu)\times \mathcal{B}(Z^{\prime}\to\uu)$. They are classified into multiplicative and additive systematic uncertainties. Their contributions are described below.

\subsection{Multiplicative systematic uncertainty}
The systematic uncertainty associated with the total number of $\jpsi$ events in data is 0.4$\%$~\cite{BESIII:2021cxx} for both processes.

The systematic uncertainty associated with the tracking efficiency is assessed using the variation in the signal efficiency caused by the track-efficiency corrections, which are applied to the $\jpsi\to\uu\uu$ MC events.  The corrections are based on the difference of the muon efficiencies 
between the data control sample and the MC sample of $\jpsi\to\gamma\uu$ in the two-dimensional plane of muon momentum $P(\mu)$ and polar angle-$\cos(\theta_{\mu})$. The changes in signal efficiency are both below $1.0\%$ for both measurements of $\mathcal{B}(\jpsi\to\uu\uu)$ and $\mathcal{B}(\jpsi\to Z^{\prime}\uu)\times \mathcal{B}(Z^{\prime}\to\uu)$. Since the resulting variation in signal efficiency from this correction is negligible, we do not apply the correction in the nominal measurement. To be conservative, we take 1.0$\%$ as the systematic uncertainty associated with the tracking for both processes.

The systematic uncertainty associated with the event selection criteria of $E_{\rm EMC}(\mu)$ and MUC depth is also assessed using the $\jpsi\to\gamma\uu$ control sample.  The difference between the data control sample and the MC sample for the two event selection criteria in the two-dimensional plane of muon momentum $P(\mu)$ and polar angle $\cos(\theta_{\mu})$ is used to re-weight the $\jpsi\to\uu\uu$ MC events, resulting in a change in signal efficiency of $-(5.0\pm0.5)\%$. Additionally, the systematic uncertainty for the $\jpsi\to Z^{\prime}(\to \uu)\uu$ process is estimated to be in the range of $(0.1-12.2)\%$ as a function of $M(Z^{\prime})$. The signal efficiencies are corrected using data for all signal MC samples, and the associated uncertainty of 0.5$\%$ is taken as the systematic uncertainty for both processes.

The systematic uncertainty associated with the kinematic fit is taken as the difference in efficiency when using and not using the helix correction \cite{BESIII:2012mpj} for the $\jpsi\to\uu\uu$ and $\jpsi\to Z^{\prime}(\to \uu)\uu$ processes. The resulting difference is 9.6$\%$ and is consistent between the two processes. We take half of this difference, 4.8$\%$, as the corresponding systematic uncertainty for both processes \cite{BESIII:2024awu}.

The systematic uncertainty associated with the simulation of the signal process $\jpsi\to\uu\uu$ is determined to be 1.0$\%$ to account for high-order effects not considered in theoretical calculations~\cite{Chen:2020bju}.

The systematic uncertainty associated with the contribution from the continuum process $\ee\to Z^{\prime}(\to\uu)\uu$ is assessed as 4.4$\%$ for the $\jpsi\to Z^{\prime}(\to \uu)\uu$ process because the cross section of $\ee\to Z^{\prime}\uu$ is only 4.4$\%$ of that for $\jpsi\to Z^{\prime}\uu$ in the $\jpsi$ data sample \cite{BESIII:2023jji}.

The uncertainty associated with the total width of the $\jpsi$~\cite{pdg}, 1.8$\%$, is taken as the systematic uncertainty for the $\jpsi\to Z^{\prime}(\to \uu)\uu$ process.

\subsection{Additive systematic uncertainties}

As described in section \ref{sec:Branching fraction and ACP measurement}, the Gaussian constraint method is applied to the yields of the various backgrounds in the fit procedure for extracting the $\jpsi\to\uu\uu$ signal yield. To estimate the systematic uncertainty associated with the background contributions, we performed an additional fit in which the background yields are fixed, and subtract the statistical uncertainty of the signal yield obtained from this extra fit from that of the standard fit in quadrature~\cite{LHCb:2013zpr}. The systematic uncertainty associated with the background contributions is determined to be 22 events, which corresponds to a 2.2\% relative uncertainty. In addition, by varying the smoothness of the signal shape, or using an alternative signal shape to improve the modeling of the high-mass tail, the uncertainty associated with signal modeling is determined to be 62 events, which corresponds to a 6.0\% relative uncertainty. For the $\jpsi\to Z^{\prime}\uu$ search, the larger of the upper limits obtained from either the fixed or Gaussian constraint background yield is adopted as the final result.

Assuming all these sources are independent, the total systematic uncertainty is obtained by adding them in quadrature.
All the sources of the systematic uncertainties are summarized in table~\ref{tab:sys_errors}.

\begin{table}[htbp]
    \centering
    \label{tab:sys_errors}
    \begin{tabular}{lcc}
    \hline
    Source & $\jpsi\to\uu\uu$ & $\jpsi\to Z^{\prime}\uu$ \\
    \hline
    \multicolumn{3}{l}{Multiplicative uncertainty (\%)} \\
    Number of $J/\psi$ events & 0.4 & 0.4 \\
    Muon tracking & 1.0 & 1.0 \\
    Muon ID requirements & 0.5 & 0.5 \\
    Kinematic fit & 4.8 & 4.8 \\
    Signal model & 1.0 & -- \\
    Continuum interference & -- & 4.4 \\
    $J/\psi$ total width & -- & 1.8 \\
    \hline
    \multicolumn{3}{l}{Additive uncertainty (\%)} \\
    Background modeling & 2.2 & -- \\
    Signal modeling & 6.0 & -- \\
    \hline
    Total (\%) & 8.1 & 6.9 \\
    \hline
    \end{tabular}
    \caption{Sources of the systematic uncertainties. The ``--'' indicates that the uncertainty is not applicable.}
\end{table}

\section{\boldmath Summary}\label{Summary}
In summary, we measure the branching fraction of the rare electromagnetic decay $\jpsi\to\uu\uu$ with improved precision and search for a new gauge boson $Z^{\prime}$ with $\jpsi\to Z^{\prime}\uu$, followed by $Z^{\prime}\to\uu$, using the world's largest $\jpsi$ data sample collected by the BESIII detector at $\sqrt{s}=3.097$~GeV. The branching fraction of $\mathcal{B}(\jpsi\to\uu\uu)$ is measured to be
\begin{equation}
\mathcal{B}(\jpsi\to\uu\uu)=[0.91\pm0.04~(\text{stat.})\pm 0.07~(\text{syst.})]\times 10^{-6},
\end{equation}
which is in good agreement with the theoretical prediction within the uncertainty, and provides a stringent test of QED predictions. We also compare the measurements of the $J/\psi\to\mu^+\mu^-\mu^+\mu^-$ branching fraction by the BESIII~\cite{BESIII:jpsi4l} (upper limit at 90$\%$ CL), CMS~\cite{CMS:2024xwi}, and LHCb~\cite{LHCb:jpsi4l} collaborations with the result from this work, as shown in figure \ref{fig:comdiffex}. It is in good agreement with other experiments but with higher precision.

The CP-odd asymmetry $A_T$ is constructed using the triple product of the independent momenta of the final-state particles in the $J/\psi$ rest frame. It is measured to be
\begin{equation}
A_T = 0.05 \pm 0.04~(\text{stat.}) \pm 0.01~(\text{syst.}),
\end{equation}
indicating that no violation of $CP$ symmetry is observed in the signal decay.

% The $CP$ asymmetry observable is constructed and measured to be:
% \begin{equation}
% A_T = 0.05 \pm 0.04~(\text{stat.}) \pm 0.01~(\text{syst.}),
% \end{equation}
% where no $CP$ violation is observed.

The $Z^{\prime}$ signal is searched for by looking for a narrow structure in the invariant mass spectrum of all possible $\uu$ combinations. Since no significant excess is observed, the upper limits on the coupling constant $g^{\prime}$ are set to be in the range 0.001 to 0.2942 for $0.212<M(Z^{\prime})<2.6$~GeV$/c^2$. BESIII provides constraints on the $g^{\prime}$ coupling comparable to those from Belle, Belle II, and BaBar within the range $M(Z^{\prime})<0.7$~GeV$/c^2$.

\begin{figure}[htp]
	\centering

	\subfigure{
		\includegraphics[width=4in,height=3in]{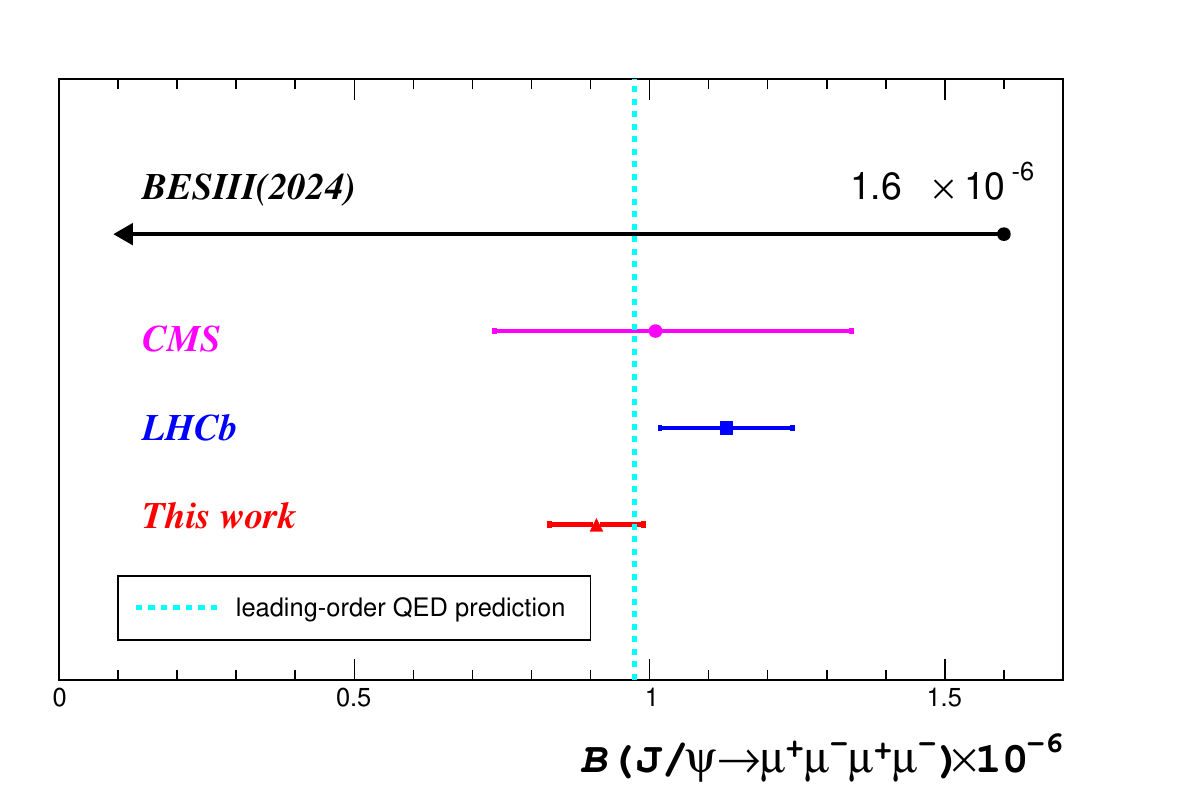}
	        }
         
	\caption{Comparison of the measured branching fractions of $J/\psi\to\mu^+\mu^-\mu^+\mu^-$ by the BESIII~\cite{BESIII:jpsi4l} (upper limit at 90$\%$ CL, $448.1 \times 10^{6}-\psi(2S) \rightarrow J/\psi \pi^{+} \pi^{-}$), CMS (33.6 ${\rm fb}^{-1}-pp \rightarrow J/\psi$)~\cite{CMS:2024xwi}, and LHCb (5.4 ${\rm fb}^{-1}-pp \rightarrow J/\psi$)~\cite{LHCb:jpsi4l} collaborations, and the result from this work (9 billion $J/\psi$). The leading-order QED prediction from ref.~\cite{Chen:2020bju} is shown as a vertical dashed line.}	
	\label{fig:comdiffex}
\end{figure}

\acknowledgments

The BESIII Collaboration thanks the staff of BEPCII (https://cstr.cn/31109.02.BEPC) and the IHEP computing center for their strong support. This work is supported in part by National Key R\&D Program of China under Contracts Nos. 2023YFA1606000, 2023YFA1606704, 2025YFA1613900; National Natural Science Foundation of China (NSFC) under Contracts Nos. 11635010, 11935015, 11935016, 11935018, 12025502, 12035009, 12035013, 12061131003, 12192260, 12192261, 12192262, 12192263, 12192264, 12192265, 12221005, 12225509, 12235017, 12342502, 12361141819, 12535005; the Chinese Academy of Sciences (CAS) Large-Scale Scientific Facility Program; the Strategic Priority Research Program of Chinese Academy of Sciences under Contract No. XDA0480600; CAS under Contract No. YSBR-101; 100 Talents Program of CAS; The Institute of Nuclear and Particle Physics (INPAC) and Shanghai Key Laboratory for Particle Physics and Cosmology; Agencia Nacional de Investigación y Desarrollo de Chile (ANID), Chile under Contract No. ANID CCTVal CIA250027; ERC under Contract No. 758462; German Research Foundation DFG under Contract No. FOR5327; Istituto Nazionale di Fisica Nucleare, Italy; Knut and Alice Wallenberg Foundation under Contracts Nos. 2021.0174, 2021.0299, 2023.0315; Ministry of Development of Turkey under Contract No. DPT2006K-120470; National Research Foundation of Korea under Contract No. RS-2026-25486791; National Science and Technology fund of Mongolia; Polish National Science Centre under Contract No. 2024/53/B/ST2/00975; STFC (United Kingdom); Swedish Research Council under Contract No. 2019.04595; U. S. Department of Energy under Contract No. DE-FG02-05ER41374.

\bibliographystyle{JHEP-mod}    \bibliography{reference}

@article{Chen:2020bju,
    author = "Chen, Wen and Jia, Yu and Mo, Zhewen and Pan, Jichen and Xiong, Xiaonu",
    title = "{Four-lepton decays of neutral vector mesons}",
    eprint = "2009.12363",
    archivePrefix = "arXiv",
    primaryClass = "hep-ph",
    doi = "10.1103/PhysRevD.104.094023",
    journal = "Phys. Rev. D",
    volume = "104",
    number = "9",
    pages = "094023",
    year = "2021"
}

@article{BESIII:jpsi4l,
    author = "Ablikim, M. and others",
    collaboration = "BESIII",
    title = "{Observation of J/{\ensuremath{\psi}} decays to $e^+e^-e^+e^-$ and $e^+e^-{\ensuremath{\mu}}^+{\ensuremath{\mu}}^-$}",
    eprint = "2111.13881",
    archivePrefix = "arXiv",
    primaryClass = "hep-ex",
    doi = "10.1103/PhysRevD.109.052006",
    journal = "Phys. Rev. D",
    volume = "109",
    number = "5",
    pages = "052006",
    year = "2024"
}

@article{pdg,
    author = "Navas, S. and others",
    collaboration = "Particle Data Group",
    title = "{Review of particle physics}",
    doi = "10.1103/PhysRevD.110.030001",
    journal = "Phys. Rev. D",
    volume = "110",
    number = "3",
    pages = "030001",
    year = "2024"
}

@article{CMS:2024xwi,
    author = "Hayrapetyan, Aram and others",
    collaboration = "CMS",
    title = "{Observation of the $J/{\ensuremath{\psi}}{\rightarrow}{\ensuremath{\mu}}^{+}{\ensuremath{\mu}}^-{\ensuremath{\mu}}^+{\ensuremath{\mu}}^-$ decay in proton-proton collisions at s=13{\,}{\,}TeV}",
    eprint = "2403.11352",
    archivePrefix = "arXiv",
    primaryClass = "hep-ex",
    reportNumber = "CMS-BPH-22-006, CERN-EP-2024-058",
    doi = "10.1103/PhysRevD.109.L111101",
    journal = "Phys. Rev. D",
    volume = "109",
    number = "11",
    pages = "L111101",
    year = "2024"
}

@article{LHCb:jpsi4l,
    author = "Aaij, Roel and others",
    collaboration = "LHCb",
    title = "{Study of the rare decay $J/\psi \to \mu^+\mu^-\mu^+\mu^-$}",
    eprint = "2408.16646",
    archivePrefix = "arXiv",
    primaryClass = "hep-ex",
    reportNumber = "LHCb-PAPER-2024-016, CERN-EP-2024-201",
    doi = "10.1007/JHEP12(2024)062",
    journal = "JHEP",
    volume = "12",
    pages = "062",
    year = "2024"
}

@article{Muong-2:2025xyk,
    author = "Aguillard, D. P. and others",
    collaboration = "Muon g-2",
    title = "{Measurement of the Positive Muon Anomalous Magnetic Moment to 127~ppb}",
    eprint = "2506.03069",
    archivePrefix = "arXiv",
    primaryClass = "hep-ex",
    reportNumber = "FERMILAB-PUB-25-0364-PPD",
    doi = "10.1103/7clf-sm2v",
    journal = "Phys. Rev. Lett.",
    volume = "135",
    number = "10",
    pages = "101802",
    year = "2025"
}

@article{Muong-2:2006E821,
    author = "Bennett, G. W. and others",
    collaboration = "Muon g-2",
    title = "{Final Report of the Muon E821 Anomalous Magnetic Moment Measurement at BNL}",
    eprint = "hep-ex/0602035",
    archivePrefix = "arXiv",
    primaryClass = "hep-ex",
    doi = "10.1103/PhysRevD.73.072003",
    journal = "Phys. Rev. D",
    volume = "73",
    pages = "072003",
    year = "2006"
}

@article{Aliberti:2025beg,
    author = "Aliberti, R. and others",
    title = "{The anomalous magnetic moment of the muon in the Standard Model: an update}",
    eprint = "2505.21476",
    archivePrefix = "arXiv",
    primaryClass = "hep-ph",
    reportNumber = "CERN-TH-2025-101, FERMILAB-PUB-25-0344-T, INT-PUB-25-015, IPARCOS-UCM-25-029, KEK Preprint 2025-22, LTH 1403, MITP-25-037, UWThPh 2025-15, UWThPh
  2025-15, ZU-TH 37/25, IPARCOS-UCM-25-029",
    doi = "10.1016/j.physrep.2025.08.002",
    journal = "Phys. Rept.",
    volume = "1143",
    pages = "1--158",
    year = "2025"
}

@article{DiLuzio:2024sps,
    author = "Di Luzio, Luca and Keshavarzi, Alexander and Masiero, Antonio and Paradisi, Paride",
    title = "{Model-Independent Tests of the Hadronic Vacuum Polarization Contribution to the Muon g-2}",
    eprint = "2408.01123",
    archivePrefix = "arXiv",
    primaryClass = "hep-ph",
    doi = "10.1103/PhysRevLett.134.011902",
    journal = "Phys. Rev. Lett.",
    volume = "134",
    number = "1",
    pages = "011902",
    year = "2025"
}

@article{BESIII:2015equ,
    author = "Ablikim, M. and others",
    collaboration = "BESIII",
    title = "{Measurement of the $e^+ e^-\to \pi^+ \pi^-$ cross section between 600 and 900 MeV using initial state radiation}",
    eprint = "1507.08188",
    archivePrefix = "arXiv",
    primaryClass = "hep-ex",
    doi = "10.1016/j.physletb.2015.11.043",
    journal = "Phys. Lett. B",
    volume = "753",
    pages = "629--638",
    year = "2016",
    note = "[Erratum: Phys.Lett.B 812, 135982 (2021)]"
}

@article{BaBar:2012bdw,
    author = "Lees, J. P. and others",
    collaboration = "BaBar",
    title = "{Precise Measurement of the $e^+ e^- \to \pi^+\pi^- (\gamma)$ Cross Section with the Initial-State Radiation Method at BABAR}",
    eprint = "1205.2228",
    archivePrefix = "arXiv",
    primaryClass = "hep-ex",
    reportNumber = "BABAR-PUB-12-003",
    doi = "10.1103/PhysRevD.86.032013",
    journal = "Phys. Rev. D",
    volume = "86",
    pages = "032013",
    year = "2012"
}

@article{CMD-2:2006gxt,
    author = "Akhmetshin, R. R. and others",
    collaboration = "CMD-2",
    title = "{High-statistics measurement of the pion form factor in the rho-meson energy range with the CMD-2 detector}",
    eprint = "hep-ex/0610021",
    archivePrefix = "arXiv",
    primaryClass = "hep-ex",
    doi = "10.1016/j.physletb.2007.01.073",
    journal = "Phys. Lett. B",
    volume = "648",
    pages = "28--38",
    year = "2007"
}

@article{Xiao:2017dqv,
    author = "Xiao, T. and Dobbs, S. and Tomaradze, A. and Seth, Kamal K. and Bonvicini, G.",
    title = "{Precision Measurement of the Hadronic Contribution to the Muon Anomalous Magnetic Moment}",
    eprint = "1712.04530",
    archivePrefix = "arXiv",
    primaryClass = "hep-ex",
    doi = "10.1103/PhysRevD.97.032012",
    journal = "Phys. Rev. D",
    volume = "97",
    number = "3",
    pages = "032012",
    year = "2018"
}

@article{He:1990pn,
    author = "He, X. G. and Joshi, Girish C. and Lew, H. and Volkas, R. R.",
    title = "{New-$Z^\prime$ phenomenology}",
    reportNumber = "UM-P-90/42, OZ-P-90/16",
    doi = "10.1103/PhysRevD.43.R22",
    journal = "Phys. Rev. D",
    volume = "43",
    pages = "22--24",
    year = "1991"
}

@article{He:1991qd,
    author = "He, Xiao-Gang and Joshi, Girish C. and Lew, H. and Volkas, R. R.",
    title = "{Simplest $Z^\prime$ model}",
    reportNumber = "CERN-TH-6084-91, UM-P-91-32, OZ-91-07",
    doi = "10.1103/PhysRevD.44.2118",
    journal = "Phys. Rev. D",
    volume = "44",
    pages = "2118--2132",
    year = "1991"
}

@article{Kamada:2018zxi,
    author = "Kamada, Ayuki and Kaneta, Kunio and Yanagi, Keisuke and Yu, Hai-Bo",
    title = "{Self-interacting dark matter and muon $g-2$ in a gauged U$(1)_{L_{\mu} - L_{\tau}}$ model}",
    eprint = "1805.00651",
    archivePrefix = "arXiv",
    primaryClass = "hep-ph",
    reportNumber = "CTPU-PTC-18-10, UMN-TH-3716/18, FTPI-MINN-18/07, UT-18-10, UMN-TH-3716-18, FTPI-MINN-18-07",
    doi = "10.1007/JHEP06(2018)117",
    journal = "JHEP",
    volume = "06",
    pages = "117",
    year = "2018"
}

@article{Foldenauer:2018zrz,
    author = "Foldenauer, Patrick",
    title = "{Light dark matter in a gauged $U(1)_{L_\mu-L_\tau}$ model}",
    eprint = "1808.03647",
    archivePrefix = "arXiv",
    primaryClass = "hep-ph",
    doi = "10.1103/PhysRevD.99.035007",
    journal = "Phys. Rev. D",
    volume = "99",
    number = "3",
    pages = "035007",
    year = "2019"
}

@article{Kahn:2018cqs,
    author = "Kahn, Yonatan and Krnjaic, Gordan and Tran, Nhan and Whitbeck, Andrew",
    title = "{M$^{3}$: a new muon missing momentum experiment to probe (g {\ensuremath{-}} 2)$_{\mu}$ and dark matter at Fermilab}",
    eprint = "1804.03144",
    archivePrefix = "arXiv",
    primaryClass = "hep-ph",
    reportNumber = "FERMILAB-PUB-18-087-A",
    doi = "10.1007/JHEP09(2018)153",
    journal = "JHEP",
    volume = "09",
    pages = "153",
    year = "2018"
}

@article{Patra:2016shz,
    author = "Patra, Sudhanwa and Rao, Soumya and Sahoo, Nirakar and Sahu, Narendra",
    title = "{Gauged $U(1)_{L_\mu - L_\tau}$ model in light of muon $g-2$ anomaly, neutrino mass and dark matter phenomenology}",
    eprint = "1607.04046",
    archivePrefix = "arXiv",
    primaryClass = "hep-ph",
    doi = "10.1016/j.nuclphysb.2017.02.010",
    journal = "Nucl. Phys. B",
    volume = "917",
    pages = "317--336",
    year = "2017"
}

@article{Shuve:2014doa,
    author = "Shuve, Brian and Yavin, Itay",
    title = "{Dark matter progenitor: Light vector boson decay into sterile neutrinos}",
    eprint = "1403.2727",
    archivePrefix = "arXiv",
    primaryClass = "hep-ph",
    doi = "10.1103/PhysRevD.89.113004",
    journal = "Phys. Rev. D",
    volume = "89",
    number = "11",
    pages = "113004",
    year = "2014"
}

@article{Langacker:2000ju,
    author = "Langacker, Paul and Plumacher, Michael",
    title = "{Flavor changing effects in theories with a heavy $Z^\prime$ boson with family nonuniversal couplings}",
    eprint = "hep-ph/0001204",
    archivePrefix = "arXiv",
    reportNumber = "UPR-0870-T",
    doi = "10.1103/PhysRevD.62.013006",
    journal = "Phys. Rev. D",
    volume = "62",
    pages = "013006",
    year = "2000"
}

@article{Chang:2009wt,
    author = "Chang, Qin and Li, Xin-Qiang and Yang, Ya-Dong",
    title = "{Constraints on the nonuniversal $Z^\prime$ couplings from B $\rightarrow$ $\pi$ K, $\pi$ K* and $\rho$ K Decays}",
    eprint = "0903.0275",
    archivePrefix = "arXiv",
    primaryClass = "hep-ph",
    doi = "10.1088/1126-6708/2009/05/056",
    journal = "JHEP",
    volume = "05",
    pages = "056",
    year = "2009"
}

@article{BaBar:2016sci,
    author = "Lees, J. P. and others",
    collaboration = "BaBar",
    title = "{Search for a muonic dark force at BABAR}",
    eprint = "1606.03501",
    archivePrefix = "arXiv",
    primaryClass = "hep-ex",
    reportNumber = "BABAR-PUB-16-003, SLAC-PUB-16549",
    doi = "10.1103/PhysRevD.94.011102",
    journal = "Phys. Rev. D",
    volume = "94",
    number = "1",
    pages = "011102",
    year = "2016"
}

@article{CMS:2018yxg,
    author = "Sirunyan, Albert M and others",
    collaboration = "CMS",
    title = "{Search for an $L_{\mu}-L_{\tau}$ gauge boson using Z$\to4\mu$ events in proton-proton collisions at $\sqrt{s} =$ 13 TeV}",
    eprint = "1808.03684",
    archivePrefix = "arXiv",
    primaryClass = "hep-ex",
    reportNumber = "CMS-EXO-18-008, CERN-EP-2018-208",
    doi = "10.1016/j.physletb.2019.01.072",
    journal = "Phys. Lett. B",
    volume = "792",
    pages = "345--368",
    year = "2019"
}

@article{Belle:2021feg,
    author = "Czank, T. and others",
    collaboration = "Belle",
    title = "{Search for Z'{\textrightarrow}${\ensuremath{\mu}}^+{\ensuremath{\mu}}^-$ in the $L_{{\ensuremath{\mu}}}-L_{{\ensuremath{\tau}}}$ gauge-symmetric model at Belle}",
    eprint = "2109.08596",
    archivePrefix = "arXiv",
    primaryClass = "hep-ex",
    reportNumber = "Belle Preprint 2021-20 KEK Preprint 2021-24",
    doi = "10.1103/PhysRevD.106.012003",
    journal = "Phys. Rev. D",
    volume = "106",
    number = "1",
    pages = "012003",
    year = "2022"
}

@article{BelleII:gp,
    author = "Adachi, I. and others",
    collaboration = "Belle-II",
    title = "{Search for a ${\ensuremath{\mu}}^+{\ensuremath{\mu}}^-$ resonance in four-muon final states at Belle II}",
    eprint = "2403.02841",
    archivePrefix = "arXiv",
    primaryClass = "hep-ex",
    reportNumber = "2024-007, 2023-55",
    doi = "10.1103/PhysRevD.109.112015",
    journal = "Phys. Rev. D",
    volume = "109",
    number = "11",
    pages = "112015",
    year = "2024"
}

@article{lum-mer,
    author = "Ablikim, M. and others",
    collaboration = "BESIII",
    title = "{Number of $J/\psi$ events at BESIII}",
    eprint = "2111.07571",
    archivePrefix = "arXiv",
    primaryClass = "hep-ex",
    doi = "10.1088/1674-1137/ac5c2e",
    journal = "Chin. Phys. C",
    volume = "46",
    number = "7",
    pages = "074001",
    year = "2022"
}

@article{Ablikim:2009aa,
    author = "Ablikim, M. and others",
    collaboration = "BESIII",
    title = "{Design and Construction of the BESIII Detector}",
    eprint = "0911.4960",
    archivePrefix = "arXiv",
    primaryClass = "physics.ins-det",
    doi = "10.1016/j.nima.2009.12.050",
    journal = "Nucl. Instrum. Meth. A",
    volume = "614",
    pages = "345--399",
    year = "2010"
}

@inproceedings{Yu:2016cof,
    author = "Yu, Chenghui and others",
    title = "{BEPCII Performance and Beam Dynamics Studies on Luminosity}",
    booktitle = "{7th International Particle Accelerator Conference}",
    doi = "10.18429/JACoW-IPAC2016-TUYA01",
    pages = "TUYA01",
    year = "2016"
}

@article{etof1,
    author = "Li, Xin and others",
    title = "{Study of MRPC technology for BESIII endcap-TOF upgrade}",
    doi = "10.1007/s41605-017-0014-2",
    journal = "Radiat. Detect. Technol. Methods",
    volume = "1",
    pages = "13",
    year = "2017"
}

@article{etof2,
    author = "Guo, Ying-Xiao and others",
    title = "{The study of time calibration for upgraded end cap TOF of BESIII}",
    doi = "10.1007/s41605-017-0012-4",
    journal = "Radiat. Detect. Technol. Methods",
    volume = "1",
    pages = "15",
    year = "2017"
}

@article{etof3,
    author = "Cao, P. and others",
    title = "{Design and construction of the new BESIII endcap Time-of-Flight system with MRPC Technology}",
    doi = "10.1016/j.nima.2019.163053",
    journal = "Nucl. Instrum. Meth. A",
    volume = "953",
    pages = "163053",
    year = "2020"
}

@article{geant4,
    author = "Agostinelli, S. and others",
    collaboration = "GEANT4",
    title = "{GEANT4 - A Simulation Toolkit}",
    reportNumber = "SLAC-PUB-9350, FERMILAB-PUB-03-339, CERN-IT-2002-003",
    doi = "10.1016/S0168-9002(03)01368-8",
    journal = "Nucl. Instrum. Meth. A",
    volume = "506",
    pages = "250--303",
    year = "2003"
}

@article{ref:kkmc1,
    author = "Jadach, S. and Ward, B. F. L. and Was, Z.",
    title = "{Coherent exclusive exponentiation for precision Monte Carlo calculations}",
    eprint = "hep-ph/0006359",
    archivePrefix = "arXiv",
    reportNumber = "CERN-TH-2000-087, UTHEP-99-09-01",
    doi = "10.1103/PhysRevD.63.113009",
    journal = "Phys. Rev. D",
    volume = "63",
    pages = "113009",
    year = "2001"
}

@article{ref:kkmc2,
    author = "Jadach, S. and Ward, B. F. L. and Was, Z.",
    title = "{The Precision Monte Carlo event generator K K for two fermion final states in $e^+e^-$ collisions}",
    eprint = "hep-ph/9912214",
    archivePrefix = "arXiv",
    reportNumber = "DESY-99-106, CERN-TH-99-235, UTHEP-99-08-01",
    doi = "10.1016/S0010-4655(00)00048-5",
    journal = "Comput. Phys. Commun.",
    volume = "130",
    pages = "260--325",
    year = "2000"
}

@article{ref:evtgen1,
    author = "Lange, D. J.",
    editor = "Erhan, S. and Schlein, P. and Rozen, Y.",
    title = "{The EvtGen particle decay simulation package}",
    doi = "10.1016/S0168-9002(01)00089-4",
    journal = "Nucl. Instrum. Meth. A",
    volume = "462",
    pages = "152--155",
    year = "2001"
}

@article{ref:evtgen2,
    author = "Ping, Rong-Gang",
    title = "{Event generators at BESIII}",
    doi = "10.1088/1674-1137/32/8/001",
    journal = "Chin. Phys. C",
    volume = "32",
    pages = "599",
    year = "2008"
}

@article{ref:lundcharm1,
    author = "Chen, J. C. and Huang, G. S. and Qi, X. R. and Zhang, D. H. and Zhu, Y. S.",
    title = "{Event generator for J/$\psi$ and $\psi$ (2S) decay}",
    doi = "10.1103/PhysRevD.62.034003",
    journal = "Phys. Rev. D",
    volume = "62",
    pages = "034003",
    year = "2000"
}

@article{ref:lundcharm2,
    author = "Yang, Rui-Ling and Ping, Rong-Gang and Chen, Hong",
    title = "{Tuning and Validation of the Lundcharm Model with $J/\psi$ Decays}",
    doi = "10.1088/0256-307X/31/6/061301",
    journal = "Chin. Phys. Lett.",
    volume = "31",
    pages = "061301",
    year = "2014"
}

@article{photos2,
    author = "Barberio, Elisabetta and van Eijk, Bob and Was, Zbigniew",
    title = "{PHOTOS: A Universal Monte Carlo for QED radiative corrections in decays}",
    reportNumber = "CERN-TH-5857-90",
    doi = "10.1016/0010-4655(91)90012-A",
    journal = "Comput. Phys. Commun.",
    volume = "66",
    pages = "115--128",
    year = "1991"
}

@article{Cvetic:2020vkk,
    author = "Cveti{\v{c}}, Gorazd and Kim, C. S. and Lee, Donghun and Sahoo, Dibyakrupa",
    title = "{Probing new physics scenarios of muon $g-2$ via $J/\psi$ decay at BESIII}",
    eprint = "2004.03124",
    archivePrefix = "arXiv",
    primaryClass = "hep-ph",
    doi = "10.1007/JHEP10(2020)207",
    journal = "JHEP",
    volume = "10",
    pages = "207",
    year = "2020"
}

@inproceedings{Rizzo:2006nw,
    author = "Rizzo, Thomas G.",
    title = "{$Z^\prime$ phenomenology and the LHC}",
    booktitle = "{Theoretical Advanced Study Institute in Elementary Particle Physics}: {Exploring New Frontiers Using Colliders and Neutrinos}",
    eprint = "hep-ph/0610104",
    archivePrefix = "arXiv",
    reportNumber = "SLAC-PUB-12129",
    pages = "537--575",
    month = "10",
    year = "2006"
}

@article{Berends:1984gf,
    author = "Berends, Frits A. and Daverveldt, P. H. and Kleiss, R.",
    title = "{Complete Lowest Order Calculations for Four Lepton Final States in electron-Positron Collisions}",
    reportNumber = "Print-84-1007 (LEIDEN)",
    doi = "10.1016/0550-3213(85)90541-3",
    journal = "Nucl. Phys. B",
    volume = "253",
    pages = "441--463",
    year = "1985"
}

@article{Rogozhnikov:2016bdp,
    author = "Rogozhnikov, A.",
    editor = "Salinas, Luis and Torres, Claudio",
    title = "{Reweighting with Boosted Decision Trees}",
    eprint = "1608.05806",
    archivePrefix = "arXiv",
    primaryClass = "physics.data-an",
    doi = "10.1088/1742-6596/762/1/012036",
    journal = "J. Phys. Conf. Ser.",
    volume = "762",
    number = "1",
    pages = "012036",
    year = "2016"
}

@article{BESIII:2021cxx,
    author = "Ablikim, M. and others",
    collaboration = "BESIII",
    title = "{Number of $J/\psi$ events at BESIII}",
    eprint = "2111.07571",
    archivePrefix = "arXiv",
    primaryClass = "hep-ex",
    doi = "10.1088/1674-1137/ac5c2e",
    journal = "Chin. Phys. C",
    volume = "46",
    number = "7",
    pages = "074001",
    year = "2022"
}

@article{Gronau:2011cf,
    author = "Gronau, Michael and Rosner, Jonathan L.",
    title = "{Triple product asymmetries in $K$, $D_{(s)}$ and $B_{(s)}$ decays}",
    eprint = "1107.1232",
    archivePrefix = "arXiv",
    primaryClass = "hep-ph",
    reportNumber = "EFI-11-16, TECHNION-PH-2011-14",
    doi = "10.1103/PhysRevD.84.096013",
    journal = "Phys. Rev. D",
    volume = "84",
    pages = "096013",
    year = "2011"
}

@article{Bigi:2017eni,
    author = "Bigi, I. I. and Kang, Xian-Wei and Li, Hai-Bo",
    title = "{CP Asymmetries in Strange Baryon Decays}",
    eprint = "1704.04708",
    archivePrefix = "arXiv",
    primaryClass = "hep-ph",
    reportNumber = "UND-HEP-17-BIG, IHEP-BESIII-LI01",
    doi = "10.1088/1674-1137/42/1/013101",
    journal = "Chin. Phys. C",
    volume = "42",
    number = "1",
    pages = "013101",
    year = "2018"
}

@article{Liu:2015uha,
    author = {Liu, Xiao-Xia and L{\"u}, Xiao-Rui and Zhu, Yong-Sheng},
    title = "{Combined estimation for multi-measurements of branching ratio}",
    eprint = "1505.01278",
    archivePrefix = "arXiv",
    primaryClass = "physics.data-an",
    doi = "10.1088/1674-1137/39/10/103001",
    journal = "Chin. Phys. C",
    volume = "39",
    number = "10",
    pages = "103001",
    year = "2015"
}

@article{Araki:2017wyg,
    author = "Araki, Takeshi and Hoshino, Shihori and Ota, Toshihiko and Sato, Joe and Shimomura, Takashi",
    title = "{Detecting the $L_{\mu}-L_{\tau}$ gauge boson at Belle II}",
    eprint = "1702.01497",
    archivePrefix = "arXiv",
    primaryClass = "hep-ph",
    reportNumber = "STUPP-17-229, UME-PP-006, YACHAY-PUB-17-01-PN",
    doi = "10.1103/PhysRevD.95.055006",
    journal = "Phys. Rev. D",
    volume = "95",
    number = "5",
    pages = "055006",
    year = "2017"
}

@article{BESIII:2023jji,
    author = "Ablikim, Medina and others",
    collaboration = "BESIII",
    title = "{Search for a muonphilic scalar $X_0$ or vector $X_1$ via J/{\ensuremath{\psi}}{\textrightarrow}{\ensuremath{\mu}}+{\ensuremath{\mu}}-+invisible decays at BESIII}",
    eprint = "2311.01076",
    archivePrefix = "arXiv",
    primaryClass = "hep-ex",
    doi = "10.1103/PhysRevD.109.L031102",
    journal = "Phys. Rev. D",
    volume = "109",
    number = "3",
    pages = "L031102",
    year = "2024"
}

@article{LHCb:2013zpr,
    author = "Aaij, R and others",
    collaboration = "LHCb",
    title = "{Measurement of $D^0--\bar D^0$ Mixing Parameters and Search for $CP$ Violation Using $D^0 \to K^+ \pi^-$ Decays}",
    eprint = "1309.6534",
    archivePrefix = "arXiv",
    primaryClass = "hep-ex",
    reportNumber = "CERN-PH-EP-2013-176, LHCB-PAPER-2013-053",
    doi = "10.1103/PhysRevLett.111.251801",
    journal = "Phys. Rev. Lett.",
    volume = "111",
    number = "25",
    pages = "251801",
    year = "2013"
}

@article{BESIII:2012mpj,
    author = "Ablikim, M. and others",
    collaboration = "BESIII",
    title = "{Search for hadronic transition $\chi_{cJ} \to \eta_c\pi^+\pi^-$ and observation of $\chi_{cJ} \to K\overline{K}\pi\pi\pi$}",
    eprint = "1208.4805",
    archivePrefix = "arXiv",
    primaryClass = "hep-ex",
    doi = "10.1103/PhysRevD.87.012002",
    journal = "Phys. Rev. D",
    volume = "87",
    number = "1",
    pages = "012002",
    year = "2013"
}

@article{BESIII:2024awu,
    author = "Ablikim, Medina and others",
    collaboration = "BESIII",
    title = "{Study of {\ensuremath{\eta}}' {\textrightarrow} {\ensuremath{\pi}}$^{+}${\ensuremath{\pi}}$^{-}$l$^{+}$l$^{-}$ decays at BESIII}",
    eprint = "2402.01993",
    archivePrefix = "arXiv",
    primaryClass = "hep-ex",
    doi = "10.1007/JHEP07(2024)135",
    journal = "JHEP",
    volume = "07",
    pages = "135",
    year = "2024"
}

@article{Aoyama:2020ynm,
    author = "Aoyama, T. and others",
    title = "{The anomalous magnetic moment of the muon in the Standard Model}",
    eprint = "2006.04822",
    archivePrefix = "arXiv",
    primaryClass = "hep-ph",
    reportNumber = "FERMILAB-PUB-20-207-T, INT-PUB-20-021, KEK Preprint 2020-5,
  MITP/20-028, KEK Preprint 2020-5, MITP/20-028, CERN-TH-2020-075, IFT-UAM/CSIC-20-74, LMU-ASC 18/20, LTH 1234,
  LU TP 20-20, LTH 1234, LU TP 20-20, MAN/HEP/2020/003, PSI-PR-20-06, UWThPh 2020-14, ZU-TH 18/20",
    doi = "10.1016/j.physrep.2020.07.006",
    journal = "Phys. Rept.",
    volume = "887",
    pages = "1--166",
    year = "2020"
}

\section*{The BESIII Collaboration}
\noindent
%% Saved at => 2026-04-17
M.~Ablikim$^{1}$\BESIIIorcid{0000-0002-3935-619X},
M.~N.~Achasov$^{4,c}$\BESIIIorcid{0000-0002-9400-8622},
P.~Adlarson$^{84}$\BESIIIorcid{0000-0001-6280-3851},
X.~C.~Ai$^{90}$\BESIIIorcid{0000-0003-3856-2415},
C.~S.~Akondi$^{32A,32B}$\BESIIIorcid{0000-0001-6303-5217},
R.~Aliberti$^{40}$\BESIIIorcid{0000-0003-3500-4012},
A.~Amoroso$^{83A,83C}$\BESIIIorcid{0000-0002-3095-8610},
Q.~An$^{79,66,\dagger}$,
Y.~H.~An$^{90}$\BESIIIorcid{0009-0008-3419-0849},
M.~S.~Anderson$^{40}$\BESIIIorcid{0009-0008-1550-2632}
% ...
\\
% \vspace{0.2cm}
% (BESIII Collaboration)\\
\vspace{0.2cm} {\it\\
$^{1}$ Institute of High Energy Physics, Beijing 100049, People's Republic of China\\
% ...
$^{4}$ Budker Institute of Nuclear Physics SB RAS (BINP), Novosibirsk 630090, Russia\\
% ...
$^{32}$ INFN Sezione di Ferrara, (A)INFN Sezione di Ferrara, I-44122, Ferrara, Italy; (B)University of Ferrara, I-44122, Ferrara, Italy\\
% ...
$^{40}$ Johannes Gutenberg University of Mainz, Johann-Joachim-Becher-Weg 45, D-55099 Mainz, Germany\\
% ...
$^{66}$ State Key Laboratory of Particle Detection and Electronics, Beijing 100049, Hefei 230026, People's Republic of China\\
% ...
$^{79}$ University of Science and Technology of China, Hefei 230026, People's Republic of China\\
% ...
$^{83}$ University of Turin and INFN, (A)University of Turin, I-10125, Turin, Italy; (B)University of Eastern Piedmont, I-15121, Alessandria, Italy; (C)INFN, I-10125, Turin, Italy\\
$^{84}$ Uppsala University, Box 516, SE-75120 Uppsala, Sweden\\
% ...
$^{90}$ Zhengzhou University, Zhengzhou 450001, People's Republic of China\\
\\
\vspace{0.2cm}
$^{\dagger}$ Deceased\\
% ...
$^{c}$ Also at the Novosibirsk State University, Novosibirsk, 630090, Russia\\
% ...
}
%% ends here %%

\end{document}